\documentclass[12pt, reqno]{amsart}
 \usepackage{amsmath, amsfonts, amssymb, amscd, mathrsfs,color, graphicx, amsthm}
\usepackage[bookmarksnumbered, colorlinks, plainpages]{hyperref}
\newtheorem{theorem}{Theorem}[section]
\newtheorem{proposition}{Proposition}[section]

\newtheorem{remark}{Remark}[section]

\makeatletter \@addtoreset{equation}{section}

\begin{document}
\title[Phase deformed magnetic Berezin transforms on $\mathbb{C}^{n}$ ]{A formula representing phase deformed magnetic Berezin transforms as functions of the magnetic Laplacian on $\mathbb{C}^{n}$}
\author[N. Askour] {Nour eddine Askour}
\address{$^\ast$ Department of Mathematics, Sultan My Slimane University, Faculty
of Sciences and Technics (M'Ghila), Beni Mellal, Morocco.}
\email{\textcolor[rgb]{0.00,0.00,0.84}{askour@fstbm.ac.ma}}
.

\begin{abstract}
we introduce a new class of a phase deformed magnetic berezin transforms and we give two formulae representing these transforms as a functions of the magnetic laplacian on $\mathbb{C}^{n}.$ As a consequence, we give an inequality of diamagnetic type.
\end{abstract} \maketitle

\section{Introduction}
The Berezin transform was introduced by Berezin \cite{bere:74} and \cite{bere:75} for certain classical domains in $\mathbb{C}$. This transform links the Berezin symbols and the symbols for Toeplitz operators. The formula represented the Berezin transform as a function of the Laplace-Beltrami operator plays an important role in the Berezin quantization theory. Classically this transform is defined as follows. consider a domain $\Omega $ $\subset \Bbb{C}^{n}$ and a Borel measure $d\mu $ on $\Omega $. Let $\mathfrak{H}$ be a closed subspace of $L^{2}\left( \Omega ,d\mu \right) $ consisting of continuous function and assume that $\mathfrak{H}$ has a reproducing kernel $K\left( .,.\right) $. The Berezin symbol $\widehat{A}$ of a bounded operator $A$ on $\mathfrak{H}$ is the function defined on $\Omega $ by

\begin{equation}
\hat{A}\left( z\right) =\frac{\left\langle AK\left( .,z\right) ,K\left( .,z\right) \right\rangle }{K\left( z,z\right) },\qquad z\in \Omega .
\end{equation}

For each $\varphi$ such that $\varphi\mathfrak{H}\in L^{2}\left( \Omega ,d\mu \right)-$
for instance for any $\varphi\in L^\infty(\Omega)$, the Toeplitz operator $T_{\varphi}$ with symbol $ \varphi$ is the operator on $\mathfrak{H}$ given by $T_\varphi[f]=P(f\varphi)$; $f\in\mathfrak{H}$, where P is the orthogonal projector on $\mathfrak{H}$.
By definition the Berezin transform B is the integral transform defined by:

\begin{equation}
 B[\varphi](z):=\widehat{T_{\varphi}}(z)=\int_{\Omega}\frac{|K(z,\omega)|^{2}}{K(z,z)}\varphi(\omega)d\mu(\omega).
 \end{equation}

Now, taking into a count that the Berezin transform can be defined provided that there is a given closed subspace, which possesses a reproducing kernel, we are here concerned with the eigenspaces

\begin{equation}
A_{m}^{2}=\{\varphi\in L^{2}(\mathbb{C},e^{-|z|^{2}}d\nu(z)), \tilde{\triangle}\varphi=E_{m}\varphi\}
\end{equation}
 of the second order differential operator

\begin{equation}
\tilde{\triangle}=-\sum_{j=1}^{n}\frac{\partial^{2}}{\partial z_{j}\partial\overline{z}_{j}}+\sum_{j=1}^{n}\overline{z}_{j}\frac{\partial}{\partial\overline{z}_{j}}
\end{equation}

 corresponding to the eigenvalues $E_{m}= m$, m=1,2,\ldots in (1,1), $d\nu$ denotes the Lebesgue measure on $\Bbb C^{n}$. These eigenspaces called generalized Bargmann spaces are reproducing kernel Hilbert space with reproducing kernels given by

\begin{equation}
K_{m}(z,w)=\frac{1}{\pi^{n}}e^{<z,w>} L_{m}^{n-1}{(|z-w|^{2})},  z,w\in\Bbb C^{n}
\end{equation}

(see \cite{AIZ:00})
It is known, that the space $A^{2}_{m}$, corresponding to $m=0$  coincides with the Bargmann-Fock space $\mathfrak{F(\mathbb{C})}$, of halomorphic function that are $e^{-|z|^{2}}d\nu$-integrable, while for $m\neq0$, The space $A_{m}^{2}$ which can be viewed as as a Kernel spaces of the Hypoelliptic  differential operator $\widetilde{\Delta}-m$, consists of non holomorphic functions. The Berezin transform $B_{m}$ is given by the following convolution operator,

\begin{equation}\label{eq:berezin}
B_{m}[\varphi](z)=\frac{m!}{\pi^{n}(n)_{m}}e^{-|w|^{2}}(L_{m}^{n-1}(|w|^{2}))^{2}\ast\varphi(z),\varphi\in L^{2}\left( \mathbb{C}^{n} ,d\nu \right)
\end{equation}

This Berezin transform can be expressed in terms of the Euclidean Laplacian as

\begin{equation}\label{eq:energyfunction0}
B_{m}=\frac{e^{\frac{{\Delta}_{\Bbb C^{n}}}{4}}}{(n)_{m}}\sum_{k=0}^{m}\frac{(n-1)_{k}(m-k)!}{k!}(\frac{\Delta_{\mathbb{C}^{n}}}{4})^{k}L_{m-k}^{k}(\frac{\Delta_{\mathbb{C}^{n}}}{4})L_{m-k}^{n-1+k}(\frac{\Delta_{\mathbb{C}^{n}}}{4})
\end{equation}

\cite{AS3:11} where $\triangle_{\Bbb C^{n}}$ the Euclidean Laplacian on $\Bbb C^{n}$ and $(\alpha)_{j})$ denotes the Pochammer symbol and $L^{(\alpha)}_{j}$ the well know Laguerre polynomials.\\
In this paper, we will be concerned with the following phase-deformed  magnetic Berezin transform,
\begin{equation}\label{eq:deformation1}
\mathfrak{B}_{m}[\varphi](z):=\int_{\mathbb{C}^{n}}e^{<z,w>}\frac{|K_{m}(z,w)|^{2}}{K_{m}(z,z)K_{m}(w,w)}\varphi(w)e^{-|w|^{2}}d\nu(w).
\end{equation}
acting on the Hilbert space $L^{2}(\Bbb{C}^{n},e^{-|w|^{2}}d\nu(w))$.
Explicitly,

\begin{equation}\label{eq:deformation2}
\mathfrak{B}_{m}[\varphi](z):=(\frac{m!}{(n)_{m}})^{2} \int_{\Bbb{C}^{n}}e^{(\left\langle z,w\right\rangle-|z-w|^{2})}(L_{m}^{n-1}(|z-w|^{2})^{2}\varphi(w)e^{-|w|^{2}}d\nu(w).
\end{equation}

Here our aim is to express the "phase deformed magnetic Berezin transform" as a function of the magnetic Laplacian $\widetilde{\Delta}$ defined by (1.4). The method used is based on the concrete $L^{2}$-spectral theory of the operator $\widetilde{\Delta}$ \cite{AS:00}, together with the functional calculus for unbounded self-adjoint operator on complex Hilbert Space \cite{birm:87},\cite{akhi:93}, Precisely we establish the following results:

\begin{equation}\label{eq:result1}
\mathfrak{B}_{m}=\pi^{n}(m!)^{2}2^{2m-n}\sum_{j=m+1}^{2m}\frac{2^{-j}}{j!(2m-j)!\Gamma(j-m+1)^{2}}\times
\end{equation}

$$
\quad_{3}F_{2}(\frac{j-2m}{2},\frac{j-2m+1}{2},j+n,j-m+1,j-m+1;1)\times
$$

$$
\frac{\Gamma(j+n+\widetilde{\Delta})}{\Gamma(n+\widetilde{\Delta})}e^{-\log(2)\widetilde{\Delta}}.
$$

\begin{equation}\label{eq:result2}
\mathfrak{B}_{m}=\frac{2^{-n}\pi^{n}}{\Gamma(n)}(\frac{m!}{(n)_{m}})^{2}\sum_{l=0}^{2m}2^{-l}\sigma^{n,m}_{l}\frac{\Gamma(n+l)}{l!}F(\widetilde{\Delta},n+l,n;\frac{1}{2}).
\end{equation}

where the coefficients $\sigma^{n,m}_{j}$ are given by:
\begin{equation}
\sigma^{n,m}_{l}=(-1)^{l}\sum_{i=0}^{l}\binom{l}{i}\binom{n+m-1}{m-l+i}\binom{n+m-1}{m-i}.
\end{equation}

This paper is organized as follows. In section 2, we recall briefly some spectral properties of the operator $\mathfrak{B}_{m},$ and as a consequence we give its spectral functions. In section 3, the first part will be devoted for the proof of the boundedness of the magnetic phase deformed Berezin transform. In the second part, we give the proof of the main results(\ref{eq:result1}) and (\ref{eq:result2}).

\section{$L^{2}$-CONCRETE SPECTRAL ANALYSIS OF THE MAGNETIC LAPLACIAN $\widetilde{\Delta}$}
In this section, we recall  some spectral properties on the $L^{2}-$ concrete spectral analysis of the magnetic Laplacian in the space of the $L^{2}-$ function on $\mathbb{C}^{n},$ with respect to the Gaussian measure \cite{AS:97}, \cite{AIZ:00}, and  \cite{AS:00}.\\
Let us fix some notations. For $p,q\in \Bbb Z_{+}$ and let $ H(p,q)$ denotes the space of restriction to the sphere $\mathcal{S}^{2n-1}=\{\omega\in \Bbb C^{n}, |\omega|=1 \}$ of the Euclidean harmonic polynomials on $ \Bbb C^{n}$ which are homogenous of degree $p$ in $z$ and degree $q$ in $ \overline{z}.$ The dimension $d(n,p,q)$ of  $ H(p,q)$ is as follows. For $n$=2,3,\ldots, we have

\begin{equation}
d(n,p,q)=\frac{(p+q-1)(p+n-2)(q+n-2)}{p!q!(n-1)!(n-2)!},
\end{equation}

 and for $n$=1, we make the convention $pq=0$; then $d(1,p,q)=1$; See \cite{Foll:75}.\\ For a complex number $\tau$, we denote by $A_{\tau}^{2}(\Bbb C^{n})$ the space of eigenfunctions $f$ of $\widetilde{\Delta}$. That is

\begin{equation}
A_{\tau}^{2}(\Bbb C^{n})=\{\psi \in L^{2}(\mathbb{C},e^{-|z|^{2}}d\nu(z)),\widetilde{\Delta}\psi=\tau \psi\}.
\end{equation}

  The concrete description of this eigenspaces is given by the following.
\textbf{proposition}\cite{AIZ:00}
Let $\tau\in \Bbb C.$ Then, we have:\\
i) For $\tau\neq0,1,2,3,...$, the space $A_{\tau}^{2}(\Bbb C^{n})$ is trivial,\\
ii)if $\tau=m\in\mathbb{Z_{+}}$
then the complex-valued function $f$ belongs to
$A_{m}^{2}(\Bbb C^{n})$
 if and only if it can be expanded in the form

\begin{equation}
f(z)=\sum_{(p,q)\in\Xi}a_{p,q}F(-m+q,n+p+q; r^{2})r^{p+q}h_{p,q}(\omega),z=r\omega,|\omega|=1,
\end{equation}

where  $F(\alpha,\gamma,x)$ is the confluent hypergeometric function,

$$
\Xi=\{(p,q)\in\mathbb{Z}\times\mathbb{Z}, p\geq0, 0\leq q\leq m\},
$$

and

$h_{p,q}=\{h_{p,q}^{j}\}_{1\leq j \leq d(n;p;q)}$ an orthonormal basis of $H(p,q)$ with  $a_{p,q}\in \Bbb C^{d(n,p,q)}$ are such that,

\begin{equation}
\sum_{(p,q)\in\Xi}\gamma(n,m;p,q)|a_{p;q}|^{2}<+\infty,
\end{equation}

where

\begin{equation}
\gamma(n,m;p,q)=\frac{(m-q)!(p+q+n-1)!}{2}\frac{\Gamma(n+p+q)}{\Gamma(m+p+q)},
\end{equation}

In \cite{AS:00}, we have proved the essentially self-adjointnees of the the operator $\widetilde{\Delta}$ with the space $C_{0}^{\infty}(\mathbb{C}^{n})$ of $\mathbb{C}-$ valued $C^{\infty}-$ function with a compact support on $\mathbb{C}^{n},$ as its natural regular domain $D(\widetilde{\Delta}),$ in $L^{2}(\mathbb{C}^{n},e^{-|z|^{2}}d\nu(z)).$ Also, the corresponding spectral family \cite{yosi:68} of $\widetilde{\Delta}$ has been given by the following proposition.

\begin{proposition}
The spectral family of the operator $\widetilde{\Delta}$ is given the following integral operator on $L^{2}(\mathbb{C},e^{-|z|^{2}}d\nu(z)):$
\begin{equation}\label{eq:spectralfamily}
E_{\lambda}[f](z)=\pi^{-n}\int_{\mathbb{C}^{n}}e^{<z,w>}L^{n}_{[\lambda]}(|z-w|^{2})f(w)e^{-|w|^{2}}d\nu(z)
\end{equation}
if $\lambda\geq0$ and $E_{\lambda}=0,$ if $\lambda<0$.
\end{proposition}
($[\lambda]$=the greatest integer not exceeding $\lambda.$)\\
For our goal, that of expressing the phase-deformed magnetic Berezin transform in termes of magnetic Laplacian $\widetilde{\Delta},$ we will need to give its spectral function \cite{esfu:99}. Precisely, we have the following proposition.

\begin{proposition}
The spectral function associated the self-adjoint operator $\widetilde{\Delta}$ is given by:
$$
e(\lambda,z,w)=(\pi)^{-n}e^{<z,w>}\sum_{m\in \Bbb Z_{+}}L_{m}^{n-1}(|z-w|^{2})\delta(\lambda-m),
$$
where $\delta(\lambda-m)$ is the Dirac delta generalized function at the point $m.$
\end{proposition}

\textbf{proof.} By a simple derivation of the equation (\ref{eq:spectralfamily}) in the distributional sense, with the use of the jumps formula (see formula (6.16) in \cite{vlad:84}, p.101), is not hard to see that the kernel (spectral function) $e(\lambda,z,w)\in\mathcal{D}'(\mathbb{R},\mathcal{D}'(\mathbb{C}^{n}\times\mathbb{C}^{n}))$ of spectral density $$\frac{dE_{\lambda}}{d\lambda}\in \mathcal{D'}(\mathbb{R},L(D(\widetilde{\Delta}),L^{2}(\mathbb{C}^{n},e^{-|z|^{2}}d\nu(z)))$$

is given by:
\begin{equation}\label{eq:spectralfunction}
e(\lambda,z,w)=(\pi)^{-n}e^{<z,w>}\sum_{k\in \Bbb Z_{+}}(L_{k}^{n}(|z-w|^{2})-L_{k-1}^{n}(|z-w|^{2}))\delta(\lambda-m)
\end{equation}

Now, using the following formula (\cite{magn:66},p241)
\begin{equation}
L_{j}^{(\alpha)}(x)=L_{j}^{(\alpha+1)}(x)-L_{j-1}^{(\alpha+1)}(x)
\end{equation}

for $j=k$ and $\alpha=n-1$, the equation (\ref{eq:spectralfunction}) becomes:

\begin{equation}
e(\lambda,z,w)=(\pi)^{-n}e^{<z,w>}\sum_{m\in \Bbb Z_{+}}L_{m}^{n-1}(|z-w|^{2})\delta(\lambda-m).
\end{equation}
This ends the proof.

\begin{remark}
For a suitable function $g:\mathbb{R}\rightarrow\mathbb{C},$ the operator $ g(\widetilde{\Delta})$ acts on $L^{2}(\Bbb C^{n},e^{-|z|^{2}}d\nu(z))$ by the following formula:
\begin{equation}\label{eq:actionoftheoperator}
g(\widetilde{\Delta})[\varphi](z)=\int_{\mathbb{C}^{n}}\Psi_{g}(z,w)\varphi(w)e^{-|w|^{2}}d\nu(w),
\end{equation}

where $\Psi_{g}(z,w)$, is the Schwartz kernel given by:

\begin{equation}\label{eq:schwartzkernel}
\Psi_{g}(z,w)= (\pi)^{-n}e^{\nu<z,w>}\sum_{m\in \Bbb Z_{+}}L_{m}^{n-1}(|z-w|^{2})g(m).
\end{equation}
Where the right hand side of the equations (\ref{eq:actionoftheoperator}) and (\ref{eq:schwartzkernel}) are understood in the distributional sense.
\end{remark}

\textbf{Remark 2.2}. Let
$$
 \widetilde{H}=\frac{-1}{4}\sum_{j=1}^{n} ((\frac{\partial}{\partial x_{j}}+iy_{j})^{2}+(\frac{\partial}{\partial y_{j}}-ix_{j})^{2}),
 $$
be the Schr\"{o}dinger operator with uniform field on $ \Bbb R^{2n},$ we can transform $\widetilde{H}$ to obtain the Magnetic Laplacian $\widetilde{\Delta}$.
 Precisely, we have:

\begin{equation}
Q\circ(\widetilde{H}-\frac{n}{2})\circ Q^{-1}=\widetilde{\Delta},
\end{equation}

Where

\begin{equation}
Qf= exp(\frac{1}{2}\sum_{j=1}^{n}(x_{j}^{2}+y_{j}^{2}))f,
\end{equation}

$f\in L^{2}(\Bbb R^{2n},dx_{1}...dx_{n}dy_{1}...dy_{n}).$

\section{phase deformed magnetic Berezin transform and the magnetic Laplacian}
In a first part of this section, we show that the phase-deformed magnetic Berezin transform is a bounded operator. The second part will be reserved to establish the two formulas expressing the phase-deformed magnetic Berezin transform as a function of the magnetic Laplacian operator.

As given in the introduction, we recall that the phase-deformed magnetic Berezin transform acting on Hilbert space $L^{2}(\mathbb{C},e^{-|z|^{2}}d\nu(z)),$ is given by:

\begin{equation}\label{eq:deformedberezin}
\mathfrak{B}_{m}[\varphi](z):=(\frac{m!}{(n)_{m}})^{2} \int_{\Bbb{C}^{n}}e^{(\left\langle z,w\right\rangle-|z-w|^{2})}(L_{m}^{n-1}(|z-w|^{2})^{2}\varphi(w)e^{-|w|^{2}}d\nu(w).
\end{equation}

We have the following proposition.

\begin{proposition}
The phase-deformed magnetic Berezin transform $\mathfrak{B}_{m},$ is a bounded operator. Precisely, we have the following estimate:

\begin{equation}
\|\mathfrak{B}_{m}[\varphi]\|_{L^{2}(\Bbb C^{n},e^{-|z|^{2}}d\nu(z))}\leq \frac{m!}{(n)_{m}}\|\varphi\|_{L^{2}(\Bbb C^{n},e^{-|z|^{2}}d\nu(z))}.
\end{equation}

\end{proposition}

\textbf{Proof}. From (\ref{eq:deformedberezin}),for $\varphi\in L^{2}(\mathbb{C},e^{-|z|^{2}}d\nu(z)),$ we have:

\begin{equation}\label{eq:inequality1}
|\mathfrak{B}_{m}[\varphi](z)|\leq (\frac{m!}{(n)_{m}})^{2}\int_{\mathbb{C}^{n}}e^{Re<z,w>-|z-w|^{2}}(L_{m}^{n-1}(|z-w|^{2})^{2}|\varphi(w)|e^{-|w|^{2}}d\nu_{\nu}(w).
\end{equation}

By using the well known inequality, $$Re(<z,w>)\leq \frac{1}{2}(|z|^{2}+|w|^{2}),$$
the inequality (\ref{eq:inequality1}) becomes:

\begin{equation}\label{eq:inequality2}
e^{\frac{-1}{2}|z|^{2}}|\mathfrak{B}_{m}[\varphi](z)|\leq (\frac{m!}{(n)_{m}})^{2}\int_{\Bbb C^{n}}e^{-|z-w|^{2}}(L_{m}^{n-1}(|z-w|^{2})^{2}e^{-\frac{1}{2}|w|^{2}}|\varphi(w)|d\nu(w).
\end{equation}

The inequality (\ref{eq:inequality2}) can be rewritten as:

\begin{equation}\label{eq:inequality3}
|e^{\frac{-1}{2}|z|^{2}}\mathfrak{B}_{m}[\varphi](z)|\leq \frac{\pi^{n}m!}{(n)_{m}}B_{m}[e^{-\frac{|.|}{2}}|\varphi|(.)](z).
\end{equation}
where, $B_{m}$ is the Berezin transform defined in (\ref{eq:berezin}).
From the inequality (\ref{eq:inequality3}), we obtain:

\begin{equation}
\|\mathfrak{B}_{m}[\varphi]\|_{L^{2}(\Bbb C^{n},e^{-|z|^{2}}d\nu(z))}\leq\frac{\pi^{n}m!}{(n)_{m}} \|B_{m}[e^{-\frac{|.|}{2}}|\varphi|(.)]\|_{L^{2}(\mathbb{C},d\nu(z))}
\end{equation}

Then, by using the following inequality see \cite{AS3:11}, p.4,
\begin{equation}
\|B_{m}[\psi]\|_{L^{2}(\mathbb{C},d\nu(z))}\leq\pi^{-n}\|\psi\|_{L^{2}(\mathbb{C},d\nu(z))}, \psi\in L^{2}(\mathbb{C},d\nu(z)),
\end{equation}

for $\psi(.)=e^{-\frac{|.|}{2}}|\varphi|(.)],$ we get:

\begin{equation}
\|\mathfrak{B}_{m}[\varphi]\|_{L^{2}(\Bbb C^{n},e^{-|z|^{2}}d\nu(z))}\leq \frac{m!}{(n)_{m}}\|\varphi\|_{L^{2}(\Bbb C^{n},e^{-|z|^{2}}d\nu(z))}.
\end{equation}

This ends the proof.

Now, we shall express the deformed Berezin transform as a function of the magnetic Laplacian on $\Bbb C^{n}.$

\begin{theorem}\label{eq:th1}
Let $m\in\Bbb Z_{+}$, then the phase deformed magnetic Berezin transform $\mathfrak{B}_{m}$ can be expressed in terms of the magnetic Laplacian

$\widetilde{\Delta}$ as

\begin{equation}
\mathfrak{B}_{m}=\frac{2^{2m-n}(m!)^{3}}{(n)_{m}}\sum_{j=m+1}^{2m}\frac{2^{-j}}{j!(2m-j)!\Gamma(j-m+1)^{2}}\times
\end{equation}

$$
\quad_{3}F_{2}(\frac{j-2m}{2},\frac{j-2m+1}{2},j+n,j-m+1,j-m+1;1)\times
$$

$$
\frac{\Gamma(j+n+\widetilde{\Delta})}{\Gamma(n+\widetilde{\Delta})}e^{-\log(2)\widetilde{\Delta}}.
$$
\end{theorem}

\textbf{Proof}.Let $g=g_{m;n}:\Bbb R\rightarrow \Bbb C,$ be a Borel function such that

\begin{equation}\label{eq:berefuncofmag}
\mathfrak{B}_{m}=g(\widetilde{\Delta}).
\end{equation}

Recalling the expression of $\mathfrak{B_{m}}$ given in (\ref{eq:deformation2}),

$$
\mathfrak{B}_{m}[\varphi](z)=(\frac{m!}{(n)_{m}})^{2} \int_{\mathbb{C}^{n}}e^{(\left\langle z,w\right\rangle-|z-w|^{2})}(L_{m}^{n-1}(|z-w|^{2})^{2}\varphi(w)e^{-|w|^{2}}d\nu(w).
$$

By using (\ref{eq:schwartzkernel}), we are lead to consider the following equality:

\begin{equation}\label{eq:disceteequation1}
(\frac{m!}{(n)_{m}})^{2}e^{<z,w>-|z-w|^{2}}(L^{n-1}_{m}(|z-w|^{2})^{2}=\pi^{-n}e^{<z,w>}\sum_{k\in \Bbb Z_{+}}L_{k}^{n-1}(|z-w|^{2})g(k).
\end{equation}

The equation (\ref{eq:disceteequation1}) can be rewritten as:

\begin{equation}\label{eq:discreteequation2}
e^{-x}(L^{n-1}_{m}(x))^{2}=\sum_{k\in\Bbb Z_{+}}h(k)L^{n-1}_{k}(x),
\end{equation}

where we have set $x=|z-w|^{2}$ and

\begin{equation}\label{eq:hrelationg}
h(k)=\pi^{-n}(\frac{(n)_{m}}{m!})^{2}g(k).
\end{equation}

By using the orthogonality relation of the Laguerre polynomials  $\{L^{n-1}_{k}\}_{k\in\mathbb{Z}_{+}}$ in the Hilbert space $L^{2}(\Bbb R,x^{n-1}e^{-x}dx),$ [\cite{niki:83}, p.37,56] the Fourier coefficient $h(k),$ is given by:

\begin{equation}
h(k)=\|L_{k}^{n-1}\|_{L^{-2}(\Bbb R,x^{n-1}e^{-x}dx)}^{-2}\int_{0}^{+\infty}x^{n-1}e^{-2x}(L_{m}^{n-1}(x))^{2}L_{k}^{n-1}(x)dx.
\end{equation}

Making use of the formula \cite{niki:83}, p.56 :
\begin{equation}\label{eq:norm}
\|L^{\alpha}_{j}\|^{2}=\frac{\Gamma(j+\alpha+1)}{j!},
\end{equation}
for $\alpha=n-1$ and $j=k,$ to obtain:
\begin{equation}\label{eq:fouriercoefficients}
h(k)=\frac{k!}{\Gamma(n+k)}\int_{0}^{+\infty}x^{n-1}e^{-2x}(L_{m}^{n-1}(x))^{2}L_{k}^{n-1}(x)dx.
\end{equation}
Now, we make use the following linearization of the product of Laguerre polynomials ([\cite{SLFD:99}, P 7361]):

\begin{equation}
L_{p}^{(\alpha}(x)L_{q}^{(\alpha}(x)=\sum_{j=|p-q|}^{p+q}l_{p,q,j}L^{(\alpha)}_{j}(x),
\end{equation}

where the coefficients are given in terms of$\quad_{3}F_{2}$ hypergeometric function [\cite{olver:74}, p.159] as

\begin{equation}
l_{p,q,j}=\frac{2^{p+q-j}p!q!}{(p+q-j)!\Gamma(j-p+1)\Gamma(j-q+1)}\times
\end{equation}

$$
\quad_{3}F_{2}(\frac{j-p-q}{2},\frac{j-p-q+1}{2},j+\alpha+1,j-p+1,j-q+1;1)
$$

 for the particular case $\alpha=n-1$ and $p=q=m.$ We obtain

\begin{equation}\label{eq:productoflaguerre}
(L^{n-1}_{m}(x))^{2}=\sum_{j=m-1}^{2m}\frac{(m!)^{2}2^{2m-j}}{(2m-j)!(\Gamma(j-m+1)^{2}}\times
\end{equation}

$$
\quad_{3}F_{2}(\frac{j-2m}{2},\frac{j-2m+1}{2},j+n,j-m+1,j-m+1;1)L^{n-1}_{m}(x).
$$

By inserting (\ref{eq:productoflaguerre}) in (\ref{eq:fouriercoefficients}), we obtain:

\begin{equation}\label{eq:foruriercoefficient1}
h(k)=\frac{k!}{\Gamma(n+k)}\sum_{j=m-1}^{2m}\frac{(m!)^{2}2^{2m-j}}{(2m-j)!(\Gamma(j-m+1)^{2}}\times
\end{equation}

$$
\quad_{3}F_{2}(\frac{j-2m}{2},\frac{j-2m+1}{2},j+n,j-m+1,j-m+1;1)\times
$$

$$
\int_{0}^{+\infty}x^{n-1}e^{-2x}L^{n-1}_{j}(x)L^{n-1}_{k}(x)dx.
$$

Next, making use of the identity [\cite{grry:07},p. 809]:

\begin{equation}
\int_{0}^{+\infty}e^{-bx}x^{\alpha}L^{(\alpha)}_{j}(\lambda x)L^{(\alpha)}_{j}(\mu x)dx=\frac{\Gamma(j+k+\alpha+1}{j!k!}\times
\end{equation}

$$
\frac{(b-\lambda)^{j}(n-\mu)^{k}}{b^{j+k+\alpha+1}}\quad_{2}F_{1}(-j,-k;-j-k-\alpha;\frac{b(b-\lambda-\mu)}{(b-\lambda)(b-\mu)})
$$

$ Re(\alpha)>-1, Re(b)>0,$ for $\alpha=n-1,$ $\lambda=\mu=1,b=2,$ the integral in (\ref{eq:foruriercoefficient1}) takes the form

\begin{equation}
\int_{0}^{+\infty}x^{n-1}e^{-2x}L^{n-1}_{j}(x)L^{n-1}_{k}(x)dx=\frac{\Gamma(j+k+n)}{j!k!2^{j+k+n}}.
\end{equation}

and (\ref{eq:fouriercoefficients}) becomes

\begin{equation}
h(k)=\sum_{j=m-1}^{2m}\frac{(m!)^{2}2^{2(m-j)-k-n}}{j!(2m-j)!(\Gamma(j-m+1))^{2}}\times
\end{equation}

$$
\quad_{3}F_{2}(\frac{j-2m}{2},\frac{j-2m+1}{2},j+n;j-m+1,j-m+1;1)\frac{\Gamma(j+k+n)}{\Gamma(k+n)}.
$$

Now, if we extend the function $h$ to the set of real numbers by:

\begin{equation}
h(\lambda)=2^{-\lambda}\sum_{j=m-1}^{2m}\frac{(m!)^{2}2^{2(m-j)-n}}{j!(2m-j)!(\Gamma(j-m+1))^{2}}\times
\end{equation}

$$
\quad_{3}F_{2}(\frac{j-2m}{2},\frac{j-2m+1}{2},j+n;j-m+1,j-m+1;1)\frac{\Gamma(j+\lambda+n)}{\Gamma(\lambda+n)}.
$$

For $\lambda\geq0$ and $h(\lambda)=0$ if $\lambda<0.$ \\ This function is well defined and satisfies the equation (\ref{eq:discreteequation2}).\\ Then, as  function $g$ satisfying (\ref{eq:hrelationg}), we can set:

\begin{equation}\label{eq:energyfunction1}
g(\lambda)=\frac{2^{2m-n}(m!)^{3}}{(n)_{m}}\sum_{j=m-1}^{2m}\frac{2^{-j}}{j!(2m-j)!(\Gamma(j-m+1))^{2}}\times
\end{equation}

$$
\quad_{3}F_{2}(\frac{j-2m}{2},\frac{j-2m+1}{2},j+n;j-m+1,j-m+1;1)\frac{\Gamma(j+\lambda+n)}{\Gamma(\lambda+n)}e^{-\log(2)\lambda};
$$

For $\lambda\geq0$ and $g(\lambda)=0$ if $\lambda<0.$\\

Now, is not hard to see that the coefficient $g(k)$ satisfies the following estimate:

\begin{equation}\label{eq:inequality12}
|g(k)|\leq\tau_{n,m}P_{n,m}(k)e^{-log(2)k}.
\end{equation}

where, $\tau_{n,m}$ is a positive constant and $P_{n,m}(k)$ is a polynomial of degree at most equal to $2m+n.$\\

On the other hand, according to the equation (7.6.11) in [\cite{szeg:39},p173], we have for a fixed $\alpha>-1$ and  $\omega$ a positive constant, the following asymptotic behavior,
\begin{equation}\label{asymptotic}
L_{k}^{\alpha}(x)=O(k^{a}), a=max(\frac{1}{2}\alpha-\frac{1}{4},\alpha),  k\rightarrow+\infty,
\end{equation}
where $x$ is a fixed number in $[0,\omega].$

Then, by the estimate (\ref{eq:inequality12}) and the asymptotic behavior (\ref{asymptotic}) considered for $\alpha=n-1,$ $x=|z-w|^{2}$ and $\omega$ be a fixed number such that $0\leq x\leq\omega,$ one can easily show that we have the following inequality:

\begin{equation}
|g(k)|L^{n-1}_{k}(|z-w|^{2})\leq\frac{Cn,m}{k^{2}},
\end{equation}
where $Cn,m$ is a positive constant and $k$ sufficiently large.

This last inequality ensures that the right hand side of the equation (\ref{eq:disceteequation1}) and the kernel $\Psi_{g}(z,w)$ given by (\ref{eq:schwartzkernel}) are well defined.\\
To close the proof of the theorem, it remains to justify that the operator $g(\widetilde{\Delta})$ given by (\ref{eq:actionoftheoperator}) is well defined as densely operator.\\

For this, let us consider a $C^{\infty}-$ function $\varphi$ with compact support in $\mathbb{C}^{n}.$

Recall that the right hand side of the equation (\ref{eq:actionoftheoperator}) is understood in the distributional sense. So, the action of the operator

$g(\widetilde{\Delta})$ on the function $\varphi$ is given by:

\begin{equation}
g(\widetilde{\Delta})[\varphi](z)=\sum_{k\in\mathbb{Z}_{+}}g(k)\int_{\mathbb{C}^{n}}e^{<z,w>}L^{n-1}_{k}(|z-w|^{2})\varphi(w)e^{-|w|^{2}}d\nu(w).
\end{equation}

We have,

\begin{equation}\label{eq:inequa0}
|\int_{\mathbb{C}^{n}}e^{<z,w>}L^{n-1}_{k}(|z-w|^{2})\varphi(w)e^{-|w|^{2}}d\nu(w)|\leq \|\varphi\|_{\infty}\int_{\mathbb{C}^{n}}e^{2Re<z,w>}|L^{n-1}_{k}(|z-w|^{2})|e^{-|w|^{2}}d\nu(w)
\end{equation}

\begin{equation}\label{eq:inequa1}
\leq\|\varphi\|_{\infty}e^{|z|^{2}}\int_{\mathbb{C}^{n}}e^{-|z-w|^{2}}|L^{n-1}_{k}(|z-w|^{2})|d\nu(w)
\end{equation}
where $\|\varphi\|_{\infty}=\sup_{w\in supp(\varphi)}|\varphi(w)|$ and $supp(\varphi)$ means the compact support of the function $\varphi.$

By using the change of variable $\xi=z-w,$ in the involved integral in the right hand side of the inequality (\ref{eq:inequa1}), we obtain:

\begin{equation}
\int_{\mathbb{C}^{n}}e^{-|z-w|^{2}}|L^{n-1}_{k}(|z-w|^{2})|d\nu(w)=\int_{\mathbb{C}^{n}}e^{-|\xi|^{2}}|L^{n-1}_{k}(|\xi|^{2})|d\nu(\xi)
\end{equation}

By applying the cauchy-Schwartz inequality to this last integral, we get:

\begin{equation}\label{eq:inequa2}
\int_{\mathbb{C}^{n}}e^{-|\xi|^{2}}|L^{n-1}_{k}(|\xi|^{2})|d\nu(\xi)\leq(\int_{\mathbb{C}^{n}}e^{-|\xi|^{2}}(L^{n-1}_{k}(|\xi|^{2}))^{2}d\nu(\xi))^{\frac{1}{2}}(\int_{\mathbb{C}^{n}}e^{-|\xi|^{2}}d\nu(\xi))^{\frac{1}{2}}.
\end{equation}

Now, let us compute the two involved integrals in the right hand side of the inequality (\ref{eq:inequa2}). For this, we use the polar coordinates $z=\rho\omega,$ $\rho>0$ and $\omega\in S^{2n-1}.$  Then, the first integral takes the following form

\begin{equation}\label{eq:equa3}
\int_{\mathbb{C}^{n}}e^{-|\xi|^{2}}(L^{n-1}_{k}(|\xi|^{2})=\Omega_{2n}\int_{0}^{+\infty}e^{-\rho^{2}}(L^{(n-1)}_{k}(\rho^{2})\rho^{2n-1}d\rho
\end{equation}

where $\Omega_{2n}=\int_{S^{2n-1}}d\sigma(\omega)=\frac{2\pi^{n}}{\Gamma(n)},$ is the area surface of the unit sphere in $\mathbb{C}^{n}.$

Next, by using the change of variable $s=\varrho^{2},$ the last integral in (\ref{eq:equa3}) becomes
\begin{equation}\label{eq:equa4}
\int_{0}^{+\infty}e^{-\rho^{2}}(L^{(n-1)}_{k}(\rho^{2})\rho^{2n-1}d\rho=\frac{1}{2}\int_{0}^{+\infty}(L^{n-1}_{k}(s))^{2}s^{n-1}e^{-s}ds.
\end{equation}

\begin{equation}
=\frac{1}{2}\|L_{k}^{n-1}\|_{L^{2}(\Bbb R,s^{n-1}e^{-s}ds)}^{2}.
\end{equation}

Finally, by using the formula (\ref{eq:norm}), for $j=k$ and $\alpha=n-1,$ the integral given in (\ref{eq:equa4}) becomes:

\begin{equation}
\int_{0}^{+\infty}e^{-\rho^{2}}(L^{(n-1)}_{k}(\rho^{2})\rho^{2n-1}d\rho=\frac{1}{2}\frac{\Gamma(k+n)}{k!}.
\end{equation}

For the second integral involved in the right hand side of the inequality (\ref{eq:inequa2}), we recognizing the well known Gaussian integral:

\begin{equation}
\int_{\mathbb{C}^{n}}e^{-|\xi|^{2}}d\nu(\xi)=\pi^{n}.
\end{equation}

Now, taking into account of the previous expressions from (\ref{eq:inequa0}), we get the following inequality:

\begin{equation}\label{eq:equa5}
|\int_{\mathbb{C}^{n}}e^{<z,w>}L^{n-1}_{k}(|z-w|^{2})\varphi(w)e^{-|w|^{2}}d\nu(w)|\leq\kappa(\varphi,n,z)\frac{\Gamma(n+k)}{k!},
\end{equation}

where $\kappa(\varphi,n,z)=\frac{2\pi^{n}}{\Gamma(n)}\|\varphi\|_{\infty}e^{|z|^{2}}.$\\
Since the right hand side of (\ref{eq:equa5}) has a polynomial growth with respect to $k$ as variable, then by using the inequality (\ref{eq:inequality12}) is not hard to show that we have the following estimate

\begin{equation}
|g(k)\int_{\mathbb{C}^{n}}e^{<z,w>}L^{n-1}_{k}(|z-w|^{2})\varphi(w)e^{-|w|^{2}}d\nu(w)|\leq\vartheta(z,\varphi,n)\frac{1}{k^{2}},
\end{equation}
where $\vartheta(z,\varphi,n)$ is a positive constant  and $k$ sufficiently large. This last inequality ensures that the right hand side of the equation (\ref{eq:actionoftheoperator}) is well defined, that is the operator $g(\widetilde{\Delta})$ is densely defined on $L^{2}(\mathbb{C},e^{-|z|^{2}}d\nu(z))$ with  $C_{0}^{\infty}(\mathbb{C}^{n})$ as its natural regular domain and satisfies the equation (\ref{eq:berefuncofmag}) . This ends the proof.

\begin{remark}

The equation (\ref{eq:discreteequation2}) is a particular case of the following identity \cite{isma:05}, p.263,

$$
e^{-ax}L^{(\alpha)}_{p}(x)e^{-ax}L^{(\alpha)}_{q}(x)e^{-ax}=\sum_{k=0}^{+\infty}c_{p,q,k}e^{-ax}L_{k}^{(\alpha)}(x),
$$
for,
 $p=q=m,$ $a=\frac{1}{2},$ and $\alpha=n-1$. This identity has been considered by Szego (1933) and Askey, and also by Gasper in connection with K.O. Frederic and H. Lewy problem.
\end{remark}

Another way to write the phase-deformed Berezin transform $\mathfrak{B}_{m},$ as a function of the magnetic Laplacian $\widetilde{\Delta}$ is as follows.

\begin{theorem}
Let $m\in\Bbb Z_{+}$, then the phase deformed magnetic Berezin transform $\mathfrak{B}_{m}$ can be expressed in terms of the magnetic Laplacian

$\widetilde{\Delta}$ as
\begin{equation}
\mathfrak{B}_{m}=\frac{2^{-n}\pi^{n}}{\Gamma(n)}(\frac{m!}{(n)_{m}})^{2}\sum_{l=0}^{2m}2^{-l}\sigma^{n,m}_{l}\frac{\Gamma(n+l)}{l!}F(-\widetilde{\Delta},n+l,n;\frac{1}{2}),
\end{equation}
with $F(a,b,c;x)$ is the Gauss hypergeometric function and the coefficients $\sigma^{n,m}_{j}$ are given by:
\begin{equation}
\sigma^{n,m}_{l}=(-1)^{l}\sum_{i=0}^{l}\binom{l}{i}\binom{n+m-1}{m-l+i}\binom{n+m-1}{m-i}.
\end{equation}
\end{theorem}
\textbf{Proof.} we return back to (\ref{eq:fouriercoefficients}) and we make use of the Feldheim formula \cite{popo:03}, which expresses the product of Laguerre polynomial as a sum of monomial terms
\begin{equation}
L^{(\alpha)}_{p}(x)L^{(\beta)}_{q}(x)=(-1)^{p+q}\sum_{l=0}^{p+q}B_{l}(p,q,\alpha,\beta)\frac{x^{l}}{l!},
\end{equation}
where the coefficient are given by:
\begin{equation}
B_{l}(p,q,\alpha,\beta)=(-1)^{p+q+l}\sum_{i=0}^{l}\binom{l}{i}\binom{\beta-q}{q-l+i}\binom{\alpha+p}{p-i},
\end{equation}
for the particular case $\alpha=\beta=n-1,$ $p=q=m.$ We obtain:
\begin{equation}
(L^{(n-1)}_{m}(x))^{2}=\sum_{l=0}^{2m}\sigma^{n,m}_{l}\frac{x^{l}}{l!}.
\end{equation}
where,
\begin{equation}
\sigma^{n,m}_{l}=(-1)^{l}\sum_{i=0}^{l}\binom{l}{i}\binom{n+m-1}{m-l+i}\binom{n+m-1}{m-i}.
\end{equation}
Therefore, equation (\ref{eq:fouriercoefficients}) takes the form
\begin{equation}\label{eq:fourier3}
h(k)=\frac{k!}{\Gamma(n+k)}\sum_{l=0}^{l=2m}\frac{\sigma^{n,m}_{l}}{l!}\int_{0}^{+\infty}x^{n+l-1}e^{-2x}L^{n-1}_{k}(x)dx.
\end{equation}

Next, making use of the identity ([\cite{grry:07},p.809]):
\begin{equation}\label{eq:specialformula}
\int_{0}^{+\infty}e^{-sx}x^{\beta}L^{(\alpha)}_{j}(x)dx=\frac{\Gamma(\beta+1)\Gamma(\alpha+j+1)}{j!\Gamma(\alpha+1)}s^{-(\beta+1)}F(-j,\beta+1;\alpha+1;\frac{1}{s}),
\end{equation}
$Re(\beta)>-1,Re(s)>0$ for $s=2, \beta=n+l-1, \alpha=n-1$ and $j=k,$ the integral in (\ref{eq:specialformula})takes the form:
\begin{equation}
\int_{0}^{+\infty}x^{n+l-1}e^{-2x}L^{n-1}_{k}(x)dx=\frac{\Gamma(n+l)\Gamma(n+k)}{k!\Gamma(n)}2^{-(n+l)}F(-k,n+l,n;\frac{1}{2})
\end{equation}

and (\ref{eq:fourier3}) becomes

\begin{equation}
h(k)=\frac{2^{-n}}{\Gamma(n)}\sum_{l=0}^{2m}2^{-l}\sigma^{n,m}_{l}\frac{\Gamma(n+l)}{l!}F(-k,n+l,n;\frac{1}{2}).
\end{equation}

Also as before, this function can be extended to the whole real line by setting:
\begin{equation}
h(\lambda)=\frac{2^{-n}}{\Gamma(n)}\sum_{l=0}^{2m}2^{-l}\sigma^{n,m}_{l}\frac{\Gamma(n+l)}{l!}F(-\lambda,n+l,n;\frac{1}{2}).
\end{equation}

if $\lambda>0$ and $h(\lambda)=0,$  for $\lambda<0.$ This function is well defined and satisfies the equation (\ref{eq:discreteequation2}).\\ Now, according to the equation (\ref{eq:hrelationg}), we can set:

\begin{equation}\label{eq:energyfunction2}
g(\lambda)=\frac{2^{-n}\pi^{n}}{\Gamma(n)}(\frac{m!}{(n)_{m}})^{2}\sum_{l=0}^{2m}2^{-l}\sigma^{n,m}_{l}\frac{\Gamma(n+l)}{l!}F(-\lambda,n+l,n;\frac{1}{2}).
\end{equation}
for $\lambda>0$ and $g(\lambda)=0,$  for $\lambda<0.$

According to (\ref{eq:hrelationg}) and (\ref{eq:fouriercoefficients}) it is natural that the function $g$ defined by (\ref{eq:energyfunction1}) and that defined by (\ref{eq:energyfunction2}) coincides on the set $\mathbb{Z}_{+}.$ Then, it follows that the rest of the proof is exactly the same as that of theorem (\ref{eq:th1}) from the inequality (\ref{eq:inequality12}). This ends the proof.

\begin{remark}
Return buck to inequality (\ref{eq:inequality3}), and replacing the phase-deformed magnetic berezin transform $\mathfrak{B}_{m}$ by its expression in terms of the magnetic laplacian $\widetilde{\Delta}$ and the magnetic Berezin transform $B_{m}$ by its expressions in terms the Euclidean Laplacian $\Delta_{C^{n}}$ then, we obtain the following inequality of Diamagnetic type:
\begin{equation}
|e^{\frac{-1}{2}|z|^{2}}g_{m}(\widetilde{\Delta})[\varphi](z)|\leq \frac{\pi^{n}m!}{(n)_{m}}f_{m}(\Delta_{C^{n}})[e^{-\frac{|.|}{2}}|\varphi|(.)](z).
\end{equation}

where $g_{m}$ is one of the functions defined by (\ref{eq:energyfunction1}) or (\ref{eq:energyfunction2}) and $f_{m}$ is the function defined by (\ref{eq:energyfunction0}).
\end{remark}

\end{document}